\documentclass[twocolumn]{aastex62}

\shorttitle{Impact of high spins on the ejection of mass in GW170817}
\shortauthors{E.R. Most et al.}

\usepackage{amsmath}

\begin{document}

\title{Impact of high spins on the ejection of mass in GW170817}

\correspondingauthor{Elias R. Most}
\email{emost@itp.uni-frankfurt.de}

\author{E.R. Most}
\affiliation{Institut f\"ur Theoretische Physik, Goethe Universit\"at
Frankfurt am Main, Germany}
\author{L.J. Papenfort}
\affiliation{Institut f\"ur Theoretische Physik, Goethe Universit\"at
Frankfurt am Main, Germany}
\author{A. Tsokaros}
\affiliation{Department of Physics, University of Illinois at
  Urbana-Champaign, Urbana, IL 61801, US}
\author{L. Rezzolla}
\affiliation{Institut f\"ur Theoretische Physik, Goethe Universit\"at
  Frankfurt am Main, Germany}



\begin{abstract}
  Following the detection of GW170817 and the accompanying kilonova
  AT2017gfo, it has become crucial to model and understand the various
  channels through which mass is ejected in neutron-star binary mergers.
  We discuss the impact that high stellar spins prior to merger have on
  the ejection of mass focussing, in particular, on the dynamically
  ejected mass by performing general-relativistic magnetohydrodynamic
  simulations employing finite-temperature equations of state and
  neutrino-cooling effects. Using eight different models with
  dimensionless spins ranging from $\chi\simeq-0.14$ to $\chi\simeq0.29$
  we discuss how the presence of different spins affects the angular
  distribution and composition of the ejected matter. Most importantly,
  we find that the dynamical component of the ejected mass can be
  strongly suppressed in the case of high spins aligned with the orbital
  angular momentum. In this case, in fact, the merger remnant has an
  excess angular momentum yielding a more extended and ``colder'' object,
  with reduced ability to shed mass dynamically. We discuss how this
  result impacts the analysis of the recent merger event GW170817 and its
  kilonova afterglow.
\end{abstract}



\section{Introduction}
\label{sec:intro}
%

Following the detection of GW170817 \citep{Abbott2017} and the subsequent
electromagnetic counterpart \citep{Abbott2017d}, it has been possible to
extract a number of different constraints and conclusions on the equation
of state (EOS) of nuclear matter. Among them are the constraints on the
maximum mass of isolated neutron nonrotating
\citep{Margalit2017,Shibata2017c,Rezzolla2017,Ruiz2017} and on the
possible ranges for radii of neutron stars
\citep{Annala2017,Most2018,Tews2018,Burgio2018,Raithel2018}. In addition
to the gravitational-wave signal,a crucial input for some of these works
is the ejected mass from the merger site that has undergone
nucleosynthesis and is hence responsible for the observed kilonova light
curves \citep{Kasen2017,Drout2017}. Hence, having a very accurate
modelling of the mass ejection and its origins is of great importance and
several studies have already been made to elucidate the ejection
mechanism and quantify the various ejection channels. Numerical
simulations classify the ejection in terms of matter that is {\it
  dynamically} ejected \citep{Hotokezaka2013, Bauswein2013b, Radice2016,
  Radice2018a, Palenzuela2015, Lehner2016, Sekiguchi2015, Sekiguchi2016,
  Dietrich2016, Dietrich2017, Dietrich2017c, Bovard2017, Papenfort2018}
during or shortly after the merger of the two stars, and in terms of
matter that is ejected {\it secularly} \citep{Siegel2017, Fernandez2018,
  Fujibayashi2017}, that is, on timescales $\gtrsim100\,{\rm ms}$. Of these two
channels, the second component is not yet very well understood, mostly
due to the lack of long-term three dimensional studies, although notable
exceptions exist, starting either from simplified initial conditions
\citep{Siegel2017, Fernandez2018} or being restricted to two spatial
dimensions \citep{Fujibayashi2017}. In comparison, the dynamically
ejected mass component has been explored in far greater detail, using
either fully consistent microphysical descriptions at finite temperature
and in full general relativity \citep{Radice2016, Radice2018a,
  Lehner2016, Sekiguchi2016, Bovard2017}, or in
approximations of general relativity \citep{Bauswein2013b}, or using a
simplified microphysics treatment \citep{Dietrich2017, Dietrich2017c,
  Hotokezaka2013, Ciolfi2017}, together with analytical expressions that
  try to combine the abundance of data available \citep{Dietrich2016}.

Another dynamically important parameter of the system influencing the
ejection of mass is the spin of the individual neutron stars. Their
magnitude or orientation are poorly known, although pulsar observations
suggest that neutron stars in binary systems can have significant spins,
such as for the binaries PSR~J1807-2500B and PSR~J1946+2052. Furthermore,
dynamical captures in globular clusters can lead to millisecond
pulsar binaries \citep{Benacquista2013}. In practice, the extraction of
information on the spin from the gravitational-wave signal of GW170817
has proved difficult so far \citep{ZhuX2018}, forcing the discussion on
the physical properties of GW170817 to be split between the ``low-'' and
``high-spin'' scenarios \citep{Abbott2017}. This uncertainty in the
modelling of the dynamical mass ejection is matched by the absence of
detailed studies for consistent spinning neutron-star merger
simulations. Studies so far have either used inconsistent initial
conditions \citep{Kastaun2013, Kastaun2017} or a simplistic model for the
description of matter \citep{Dietrich2017c}. More specifically, although
\cite{Dietrich2017c} have studied in great detail the effect of spin on
the gravitational-wave signal and on the mass ejection, only small
dimensionless spins $\chi\simeq0.1$ were used. Furthermore, the absence
of neutrino interactions makes it difficult to classify the reasonable
amount of shock heating and composition of the ejected matter.

In this Letter we attempt to fill this gap and study the mass ejection of
high spin systems up to $\chi\simeq0.29$ for two finite-temperature EOSs,
representing both high and low compactness, with the latter being
favoured by the detection \citep{Annala2017, Most2018, Abbott2018b}.

\begin{table}
  \centering
  \setlength{\tabcolsep}{0.25em}
  \begin{tabular}{|c|c|c|c|c|c|}
    \hline
    $\chi$ & $\Omega\times 10^{-3}$     &  $J_{_{\rm ADM}}$           & $P$                       & $M_{\rm ej} \times 10^{-3}$ & ${\rm EOS}$ \\
           & $\left[M_\odot^{-1}\right]$ &  $\left[M_\odot^2\right]$ & $\left[{\rm ms}\right]$   & $\left[M_\odot\right]$    &             \\
    \hline
    $-0.148\ (\downarrow\downarrow)$        & $8.77$ & $6.89$ & $-3.34$           & $4.41$     & \footnotesize{$\rm TNTYST$} \\
    $-0.002\ (00)$                          & $8.75$ & $7.37$ & $-258$            & $2.73$     & \footnotesize{$\rm TNTYST$} \\
    $\phantom{-}0.106\ \qquad$              & $8.74$ & $7.72$ & $\phantom{-}4.49$ & $0.31$     & \footnotesize{$\rm TNTYST$} \\
    $\phantom{-}0.287 \ (\uparrow\uparrow)$ & $8.81$ & $8.39$ & $\phantom{-}1.75$ & $0.24$     & \footnotesize{$\rm TNTYST$} \\ 
    \hline                                                                                            
    $-0.142\ (\downarrow\downarrow)$        & $8.76$ & $6.90$ & $-4.35$           & $1.45$     & \footnotesize{${\rm BHB}\Lambda\Phi$} \\
    $-0.001\ (00)$                          & $8.75$ & $7.37$ & $-619$            & $0.64$     & \footnotesize{${\rm BHB}\Lambda\Phi$} \\
    $\phantom{-}0.156\qquad\ $              & $8.75$ & $7.90$ & $\phantom{-}3.93$ & $0.62$     & \footnotesize{${\rm BHB}\Lambda\Phi$} \\
    $\phantom{-}0.194\ (\uparrow\uparrow)$  & $8.75$ & $8.04$ & $\phantom{-}3.18$ & $0.37$     & \footnotesize{${\rm BHB}\Lambda\Phi$} \\
    \hline
  \end{tabular}
  \caption{Summary of the binaries considered. Reported are: the
    dimensionless spin $\chi$, the period $P$, the orbital angular
    frequency $\Omega$, the ADM angular momentum $J_{_{\rm ADM}}$, and
    the ejected mass $M_{\rm ej}$. All binaries have $M_{_{\rm
        ADM}}=2.700$ and are at an initial separation of $45\,{\rm
      km}$. The labels $(\downarrow\downarrow)$, $(00)$ and
    $(\uparrow\uparrow)$ refer to reference binaries.}\label{tab:ID}
\end{table}

\section{Methods}
\label{sec:methods}
\begin{figure}[t]
  \centering
  \includegraphics[width=0.95\columnwidth]{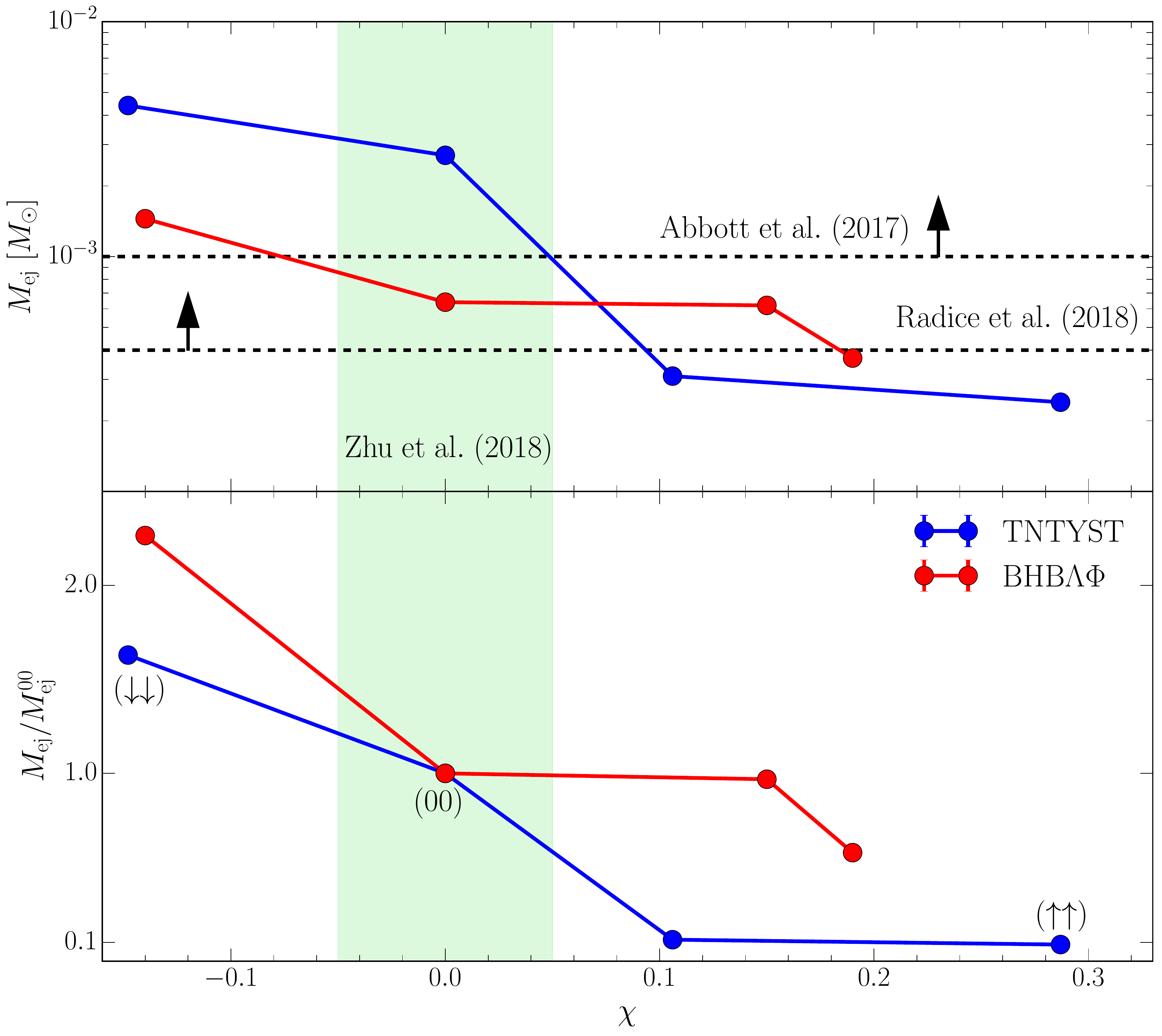}
  \caption{Dynamically ejected matter $M_{\rm ej}$ as a function of the
    dimensionless spin $\chi$ of the stars in the binary. The upper panel
    shows the absolute value of the ejected mass, while the lower panel
    the relative value normalised to the irrotational case $M^{00}_{\rm
      ej}$. The dashed lines refer to lower-bound values in the
    literature.}
  \label{fig:Mej}
\end{figure}

This work studies the merger and early post-merger phase of binary
neutron-star coalescence with non-vanishing spins. Spinning BNS in
quasi-equilibrium are modeled via the \texttt{COCAL} code
\citep{Tsokaros2015,Tsokaros2018}, which uses three spherical grids
together with a second-order finite difference scheme to invert the
elliptic equations of the gravitational and fluid potentials. The
formulation used to obtain the spinning configurations is a modification
of the formulation of irrotational binaries where the fluid velocity is
decomposed into an irrotational and a spinning part \citep{Tichy11}.  The
latter is an input quantity that controls the final dimensionless spin
$\chi$ that is quoted in this work. Here we define $\chi:=J_{_{\rm
    QL}}/(M_{_{\rm ADM}}/2)^2$, where $J_{_{\rm QL}}$ is the quasi-local
angular momentum of each neutron star, and $M_{_{\rm ADM}}$ the
Arnowitt-Deser-Misner mass of the system \citep{Tsokaros2018}. In Table
\ref{tab:ID} we give an overview of the binaries studied in this work,
whose components both have the same mass and a total $M_{_{\rm
    ADM}}=2.700\,M_{\odot}$. Since producing accurate initial data with
high spins is computationally very demanding, we focus on only two spin
configurations, either aligned $(\uparrow\uparrow)$ or misaligned
$(\downarrow\downarrow)$ with the orbital angular momentum, but consider
rotation periods of up to $1.7\,{\rm ms}$, in line with millisecond
pulsars like PSR J1748-2446ad \citep{Hessels2006}.

\begin{figure*}[ht]
\begin{center}
  \centering
  \includegraphics[width=0.44\textwidth, trim=0 0 60 0, clip]{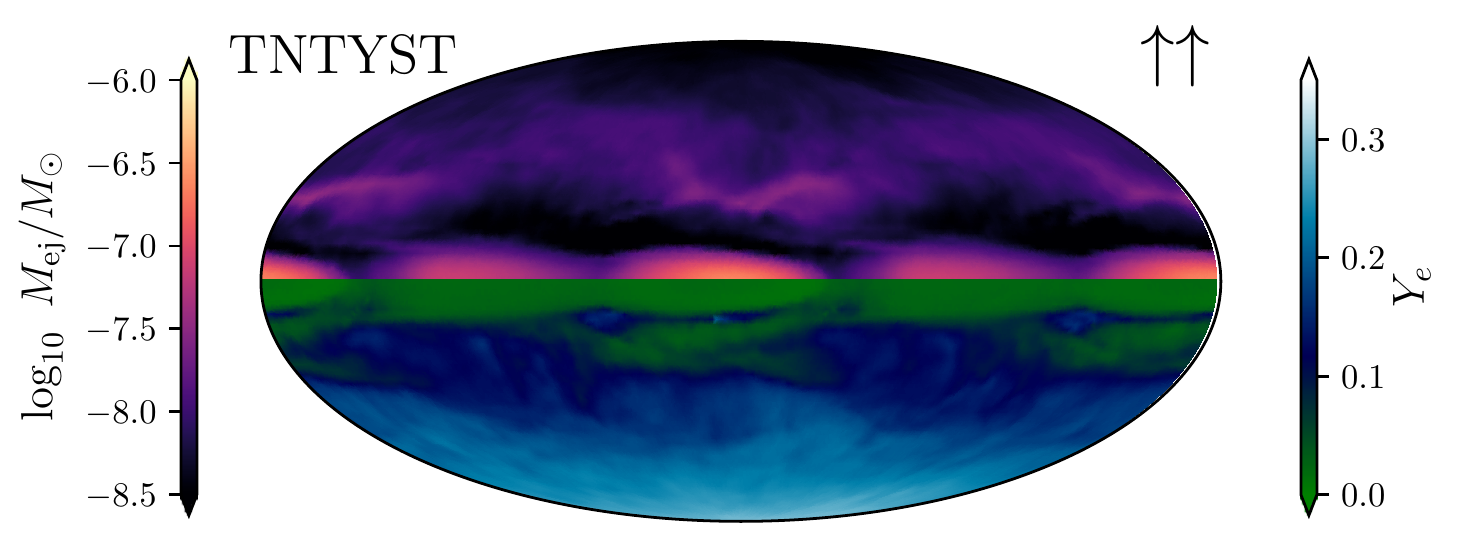}
  \includegraphics[width=0.44\textwidth, trim=60 0 0 0, clip]{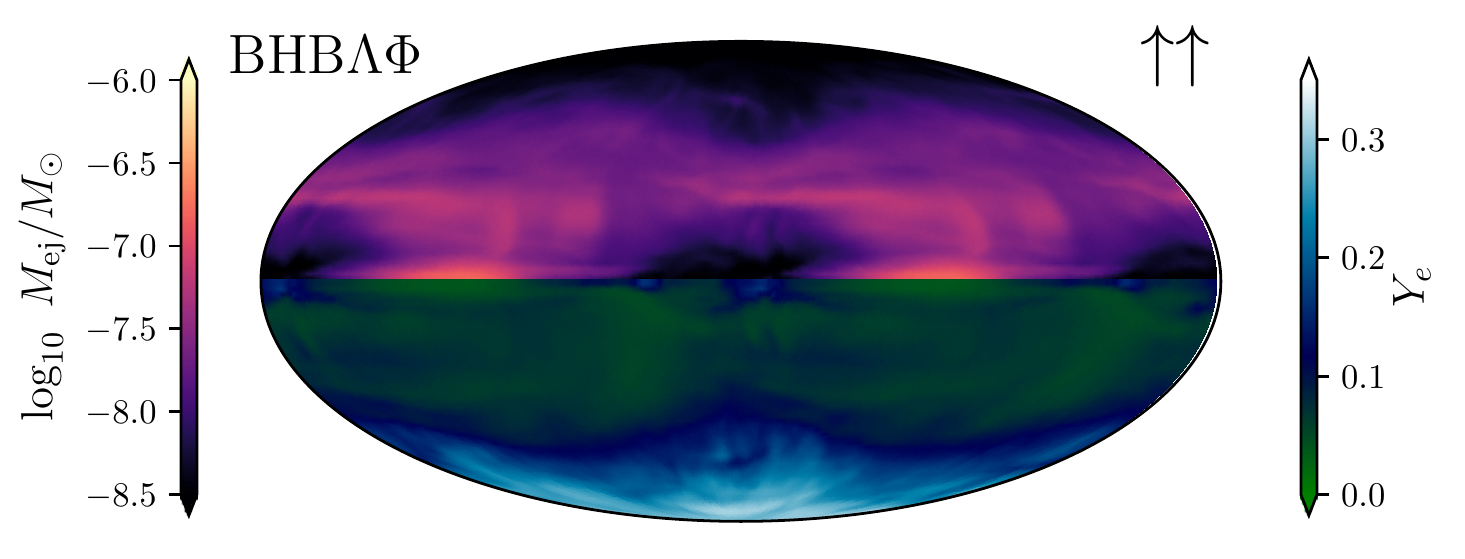}
  \\
  \includegraphics[width=0.44\textwidth, trim=0 0 60 0, clip]{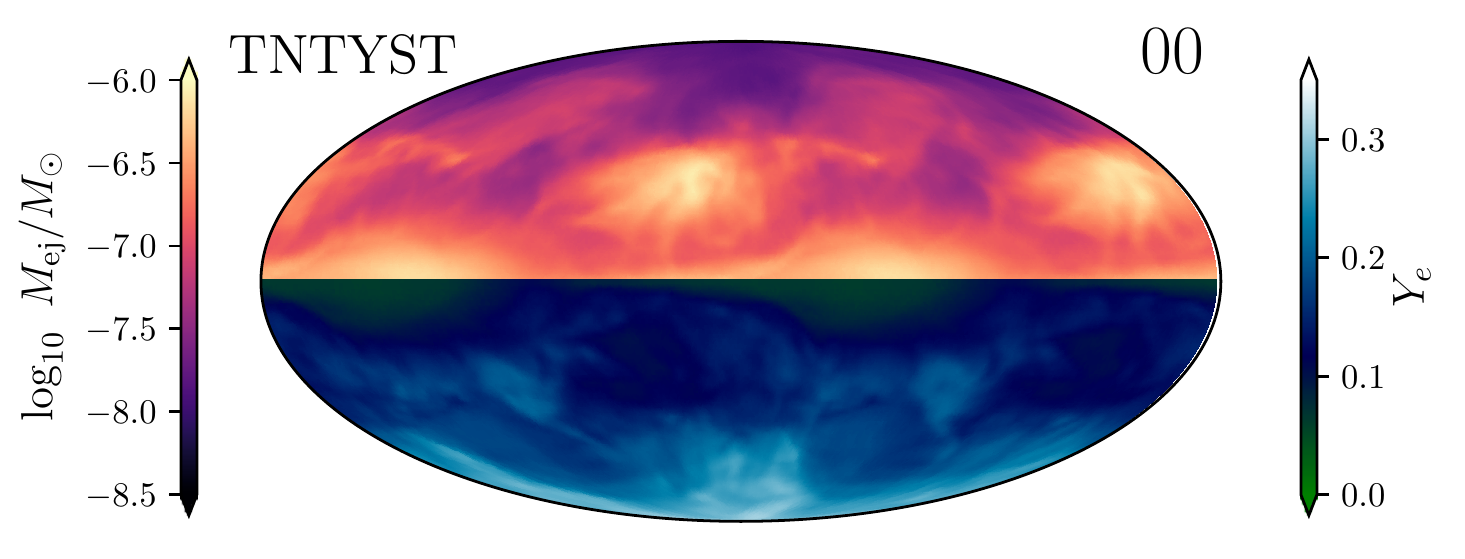}
  \includegraphics[width=0.44\textwidth, trim=60 0 0 0, clip]{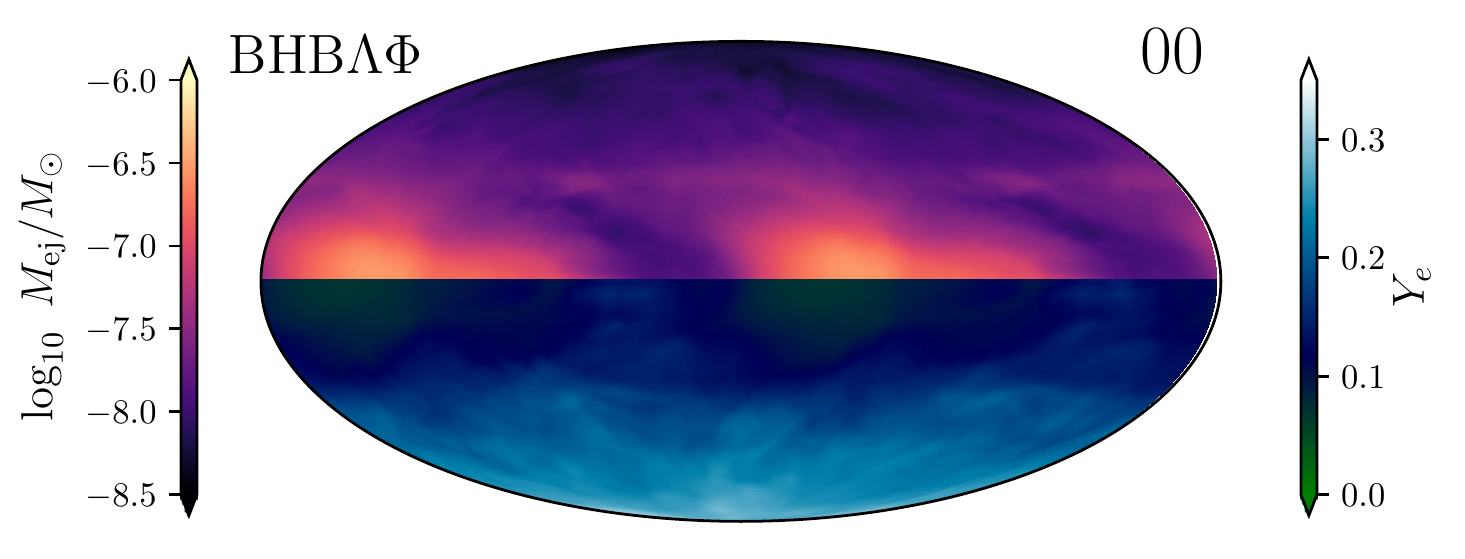}
  \\
  \includegraphics[width=0.44\textwidth, trim=0 0 60 0, clip]{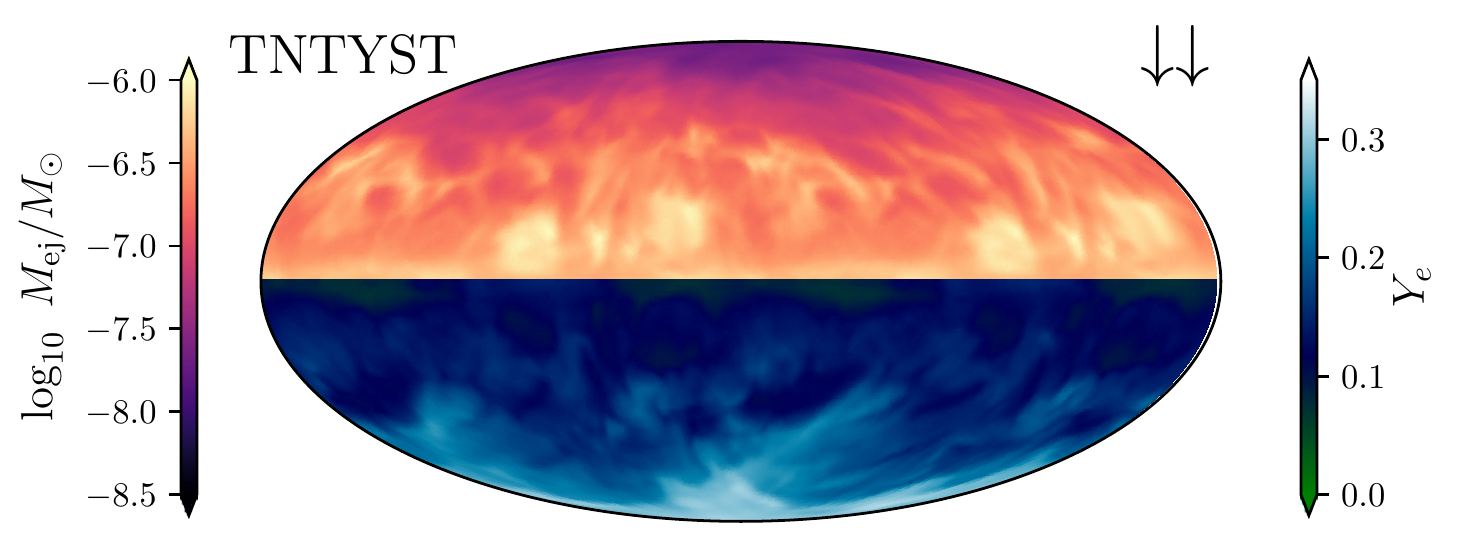}
  \includegraphics[width=0.44\textwidth, trim=60 0 0 0, clip]{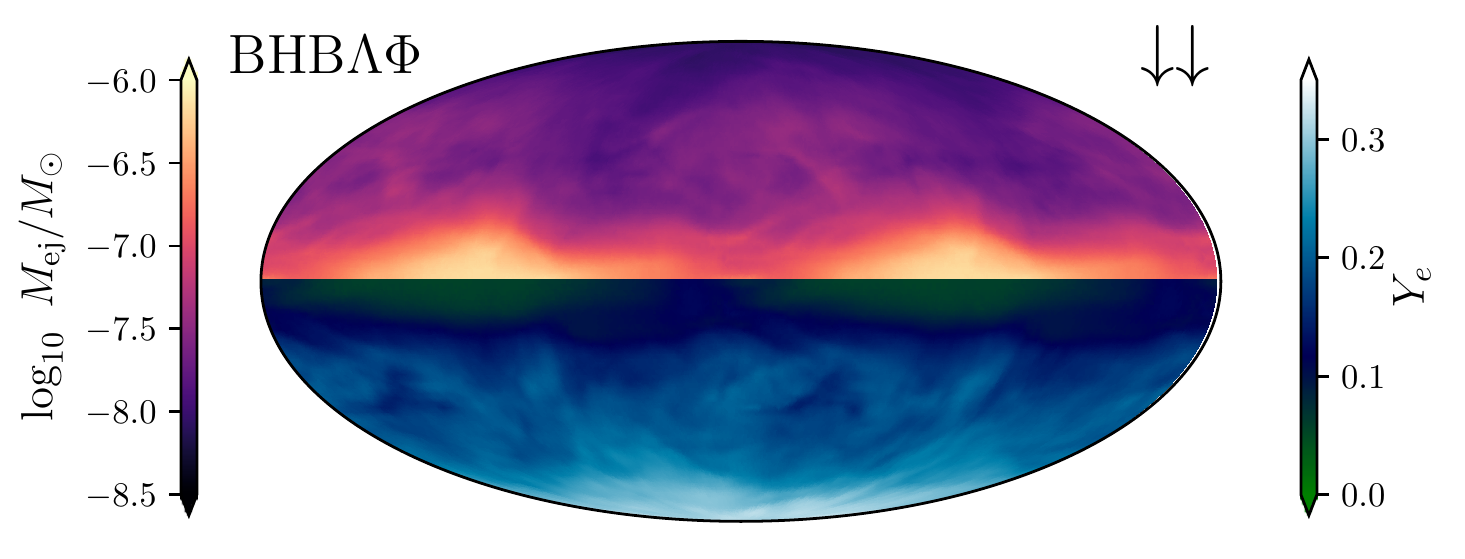}
  \caption{Angular distribution of the time-integrated ejected mass
    $M_{\rm ej}$ and average electron fraction $Y_e$ for systems using
    the $\rm TNTYST$ \textit{(left column)} and ${\rm BHB}\Lambda\Phi$
    \textit{(left column)} EOS.}
  \label{fig:Mej_2D}
\end{center}
\end{figure*}

We solve the coupled Einstein--general-relativistic magnetohydrodynamic
(GRMHD) system using the \texttt{Frankfurt\-/IllinoisGRMHD} code
(\texttt{FIL}) \citep{Most2018b}, which is derived from the
\texttt{IllinoisGRMHD} code \citep{Etienne2015}. The GRMHD equations are
solved using fourth-order accurate conservative finite differencing.  To
solve the Einstein equations we implement a fourth-order accurate
discretization of the Z4c system \citep{Hilditch2012}.  Neutrino cooling
and weak interactions are incorporated via a leakage scheme
\cite{Galeazzi2013}. The codes makes use of the publicly available
Einstein Toolkit framework \citep{loeffler_2011_et}. The computational
domain is given by a set of nested boxes in a fixed mesh refinement
setting extending up to $\simeq 1500\,{\rm km}$ and using a total of
seven levels with a highest resolution of $\simeq 250\,{\rm m}$. To
assess the impact of resolution, we also performed simulations for three
$\rm TNTYST$ models at $\simeq 370\,{\rm m}$ and one for the irrotational
system at $\simeq 195\,{\rm m}$. In all cases, the relative difference
were between $10\%$ and $20\%$, thus much less than the uncertainty
stemming from the criterion for unboundness \citep{Bovard2017}.

The final important ingredient is the description of nuclear matter at
finite temperatures. In light of recent studies following the detection
of GW170817 \citep{Annala2017, Most2018, Abbott2018b}, we select two
temperature-dependent EOSs: $\rm TNTYST$ \citep{Togashi2017} and ${\rm
  BHB}\Lambda\Phi$ \citep{Banik2014}, to representing the bounds on small
and high tidal deformabilities.

Every star in the binary systems is initially endowed with a poloidal
magnetic field of $10^{15}\,{\rm G}$ at its center. While magnetic fields
are very important to study secular outflows, the impact they have on the
dynamical mass ejection are small (for the $\rm TNTYST$ irrotational
binary, the magnetically induced differences in the mass ejection rates
are $\lesssim5\,\%$) and we will not discuss them here for compactness.

\section{Results}
\label{sec:results}

We considered a total of eight systems modeled using two EOSs and having
either spins aligned or misaligned with the orbital spin angular
momentum, ignoring systems where only one of the stars is spinning
\citep{Dietrich2017c}. While we plan to consider these effects in future
work, we note that unequal-mass binaries are expected to yield
systematically larger matter outflows~\citep{Rezzolla:2010,Lehner2016,
  Sekiguchi2016}, although for mass ratios ($q\gtrsim0.8$),
  this effect is not expected to be significant \citep{Dietrich2017c}.

Depending on the spin orientation and starting from a separation of
$45\,{\rm km}$, the two stars inspiral for several orbits before they
merge. As first shown by \citealt{Kastaun2013}, systems with large
aligned spins take longer to radiate away the orbital angular momentum
and will hence merge later than the systems with misaligned spin
\citep[see also][for a detailed description]{Dietrich2017c,
  Ruiz2019}. After the merger, we record the mass that is unbound
according to the geodesic criterion \citep{Bovard2017} and crosses a
sphere at a radius of $600\,{\rm km}$, thus obtaining the dynamically
ejected mass by an integration over roughly $15-20\,{\rm ms}$, when the
mass flux drops significantly in all cases. We note that all simulations
were carried out for sufficiently long timescales to observe a vanishing
dynamical ejection of matter and that no black hole was formed by the end of
the simulations.

Before focussing on three fiducial cases -- misaligned high spins
($\downarrow\downarrow$), no spins ($00$), and aligned high spins
($\uparrow\uparrow$) -- we give an overview of the amount of dynamically
ejected matter for all models listed in Table \ref{tab:ID}. This is shown
in Fig. \ref{fig:Mej}, which reports the amount of dynamically ejected
mass depending on the dimensionless spin $\chi$. The upper panel refers
to the absolute value of the ejected mass, while the lower panel
emphasizes the difference with the respect to the irrotational case,
which is the most common in merger simulations.

Note that intermediate misaligned spins $\chi\sim-0.14$ lead only to a
small twofold increase compared with the irrotational case, while for
high aligned spins $\chi\lesssim0.29$ we find that large aligned spins
can significantly reduce the mass ejection by about a factor of two in
the case of the stiff ${\rm BHB}\Lambda\Phi$ EOS and even of about one
order of magnitude in the case of the soft $\rm TNTYST$ EOS. As a
comparison, we also show in the top panel of Fig. \ref{fig:Mej} the
reference bounds on the dynamical mass ejection coming either from the
LIGO GW170817 analysis \citep{Abbott2017c} or by considering the models
having $M\geq2.7\,M_\odot$ in \citet{Radice2018a}. Clearly, most of our
aligned spinning cases produce lower values than expected from these
studies and this contrast in the dynamical ejecta can only become larger
with unequal-mass binaries since the secondary star is tidally disrupted
\citep{Rezzolla:2010}. We also note that we have tried to use the fitting
formulas for the dynamically ejected mass provided by
\citet{Dietrich2016} and calibrated by \citet{Radice2018a}, finding
negative values when considering the $\rm TNTYST$ EOS. This highlights
that more simulations are needed to achieve a comprehensive -- and thus
accurate -- coverage of the space of EOS and spin configurations.

\begin{figure}[t]
  \centering
  \includegraphics[width=0.45\textwidth]{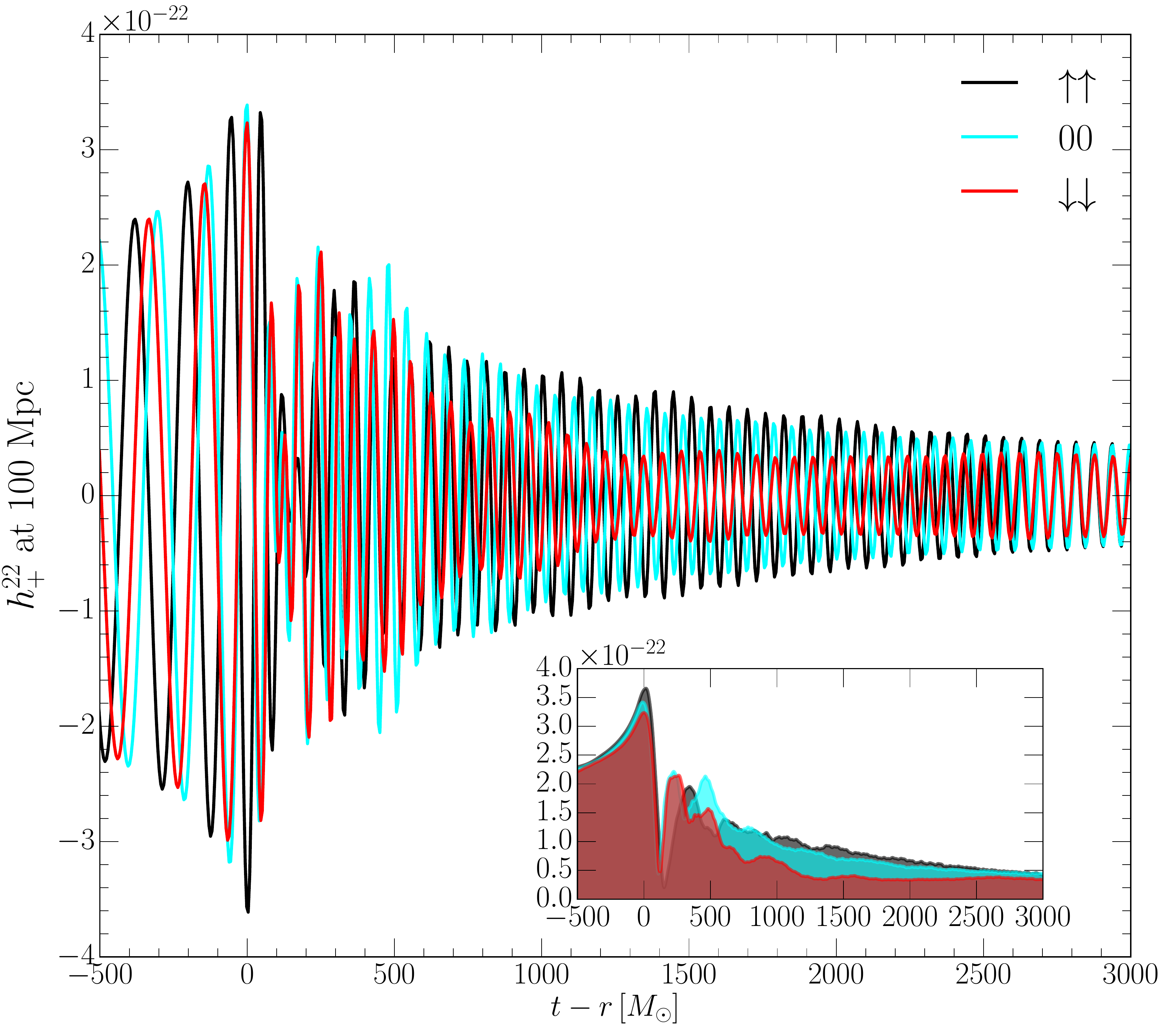}
  \caption{Gravitational-wave strains ($\ell=2=m$ mode of the $h_{+}$
    polarization) for binaries with the TNTYST EOS; the amplitudes of the
    signals are shown in the inset.}
  \label{fig:GW}
\end{figure}

\begin{figure*}[t!]
\begin{center}
  \centering
  \includegraphics[width=0.9\textwidth]{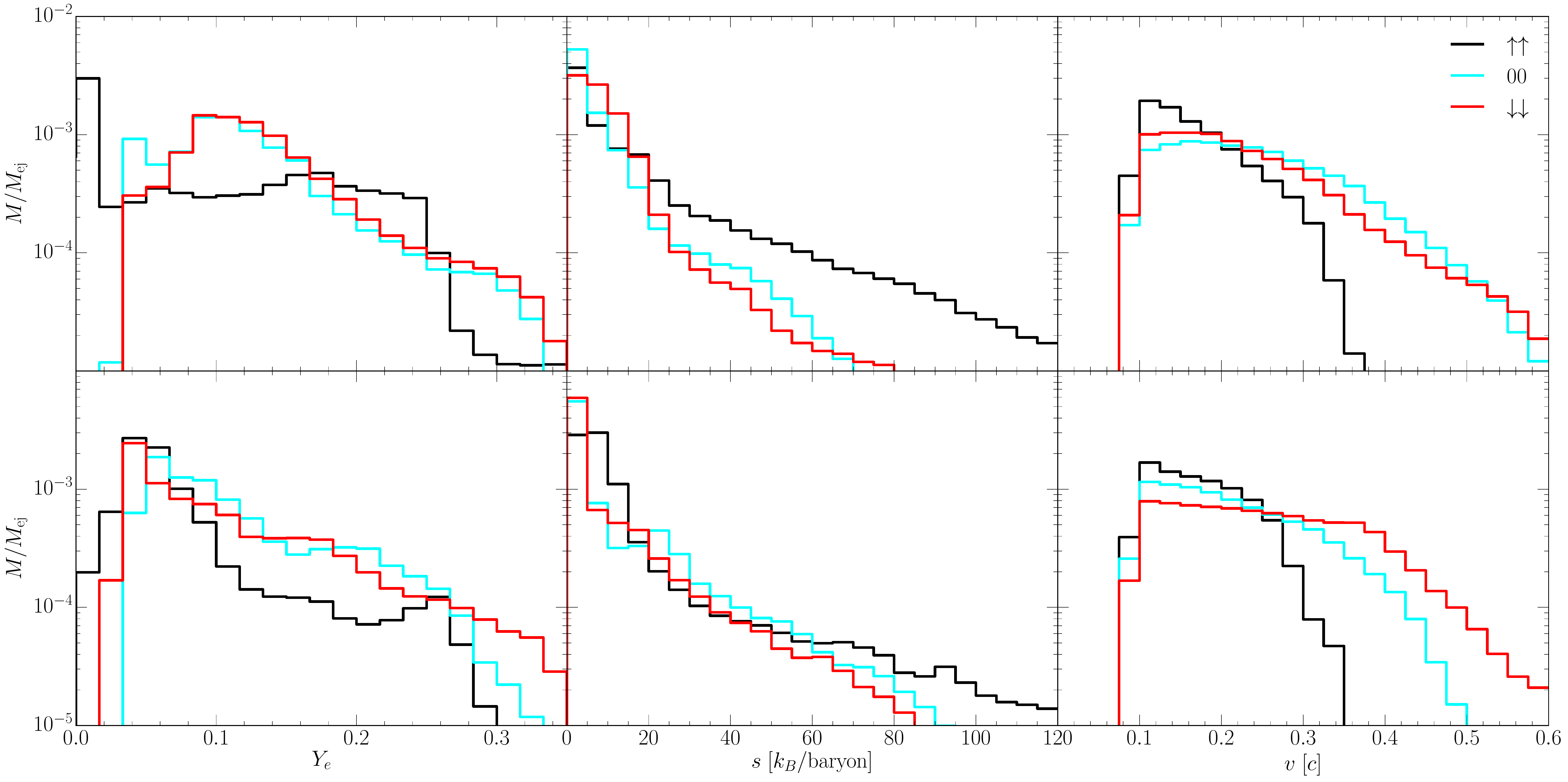}
  \caption{Histograms showing various composition features of the ejected
    mass. From left to right: the electron fraction $Y_e$, the entropy
    $s$ per baryon and the velocity $v$ as computed from the Lorentz
    factor of the outflow. The {\it top row} refers to models using the
    $\rm TNTYST$ EOS, while the {\it bottom row} to the ${\rm
      BHB}\Lambda\Phi$ models.}
  \label{fig:hist}
\end{center}
\end{figure*}

To better represent the drastic change in mass ejection and also to
elucidate its properties with regards to an accompanying kilonova, we
show in Fig. \ref{fig:Mej_2D} the angular distributions of the ejected
mass projected onto a two-sphere at $600\,{\rm km}$. Starting with the
irrotational case (middle row in Fig. \ref{fig:Mej_2D}), the top part of
the projection illustrates that ejected mass is mainly distributed along
the equatorial plane as a result of tidally driven outflow (see also
\citealt{Sekiguchi2016, Bovard2017, Radice2018a} for a more detailed
discussion). The behaviour is rather similar for the two EOSs considered,
but in the case of the softer $\rm TNTYST$ EOS, shock-heating also plays
a role in the ejection of matter, which is visible at a latitude of
$\sim45$ degrees. Additionally, the part of the same panel reportes the
time-averaged composition of the ejected matter, pointing out that while
the tidal component is very neutron rich, i.e., $Y_e\lesssim0.1$, there
are regions of high protonization at higher latitudes and conversely
lower densities extending the $Y_e$ range to $\lesssim 0.3$.

When contrasting the irrotational binaries with the misaligned-spin ones
(bottom row in Fig. \ref{fig:Mej_2D}), it is clear that in this case most
of the ejecta is driven by strong shock-heating -- especially in the case
of the $\rm TNTYST$ EOS -- and are distributed to higher latitudes. This
is due to the fact that the misaligned binaries not only have the
smallest amount of angular momentum after merger, and so higher radial
velocities in the ejecta (see also Fig. \ref{fig:hist}), but also the
largest plunge velocity at merger. Indeed, these binaries inspiral with
fewer orbits and merge with a radial velocity for the $\rm TNTYST$ (${\rm
  BHB}\Lambda\Phi$) EOS that -- in units of the speed of light -- is
$\simeq0.042\,(0.035)$; these plunge velocities should be compared with
$\simeq0.028\,(0.027)$ for the aligned case.  On the other hand, when
considering aligned high-spin binaries (top row in
Fig. \ref{fig:Mej_2D}), we find that mass ejection is significantly
suppressed, up to one order of magnitude for the $\rm TNTYST$
EOS. Furthermore, the ejection is strongly beamed towards the equator, a
clear indication that the origin of the mass ejection is purely tidal,
consistent with the fact of the two stars having a significantly reduced
relative velocity at merger, in analogy with what happens in eccentric
mergers \citep{Radice2016, Papenfort2018}. An important difference with
respect to the other two cases is that the ejecta are mainly neutron rich
and even the polar regions feature significantly smaller electron
fractions.

Much of the phenomenology discussed above is reflected in the
gravitational-wave emission from the merger remnant. Figure \ref{fig:GW}
reports the amplitude of gravitational-wave strain of the three reference
cases and it is apparent that the remnant from the binary with large
aligned spins is less compact because of the excess angular momentum;
this more extended and ``colder'' object will produce comparatively
larger-amplitude gravitational waves and a reduced ejection of
neutron-rich matter. By contrast, the remnant from the binary with large
misaligned spins is more compact and ``hotter'', with a reduced emission
of gravitational waves, but a larger ejection of high-$Y_e$ matter. As
expected, the irrotational binary is intermediate between these two
cases.

Finally, we report in Fig. \ref{fig:hist} the relative distributions of
$Y_e$, of the entropy $s$ per baryon and of the velocity $v$ of the
ejecta. For both EOSs, the behaviour is qualitatively the same: the
misaligned and the irrotational binaries have similar compositions
peaking around $Y_e\simeq0.1$ and then rapidly falling off until about
$Y_e\simeq0.35$. Most of these ejecta have small entropies
$s<20\,k_B/{\rm baryon}$, with $s<80\,k_B/{\rm baryon}$ almost
everywhere. The material for these two systems also has outflow
velocities reaching up to $0.6$. On the other hand, the aligned spinning
binaries have large amounts of ejecta around $Y_e < 0.05$ and almost no
material is present with $Y_e > 0.3$; a cut-off in the electron fraction
is present in both EOSs. Similarly, the velocity $v$ of the ejecta in the
aligned binaries peaks around small values $v\simeq0.1$, pointing to the
absence of strong shock heating in most of the matter, with a cut-off
velocity of $v\simeq0.35$ for both EOSs.

\section{Conclusions}
\label{sec:conclusions}

We have presented GRMHD simulations of neutron-star mergers having high
component spins and employing two temperature-dependent EOSs representing
the soft and stiff limits of current constraints \citep{Annala2017,
  Most2018, Abbott2018b}. By varying the orientation as well as the
magnitude of both stellar spins, we were able to reveal a strong impact
on dynamical mass ejection and highlight the changes in the nuclear
composition of the material. In particular, binaries with high spins
aligned with the orbital angular momentum lead to a strong suppression of
the ejected mass when compared to the standard case of irrotational
binaries. At the same time, misaligned configurations show only a modest
increase. We found that this effect was most pronounced for the soft $\rm
TNTYST$ EOS, where it lead to a suppression of one order of magnitude
compared with the non-spinning case.

The physical origin of this behaviour lies with an increased total
angular momentum at merger of the binaries with aligned spins and
consequently with the decreased plunge velocities when the two stars
collide. In turn, this process leads to a merger remnant that is more
extended and ``colder'' than in the case of irrotational binaries, thus
with a reduced ability to shed mass dynamically, but more efficient in
radiating gravitational waves. In contrast, binaries with large
misaligned spins collide with larger velocities and yield, for the same
EOS, a remnant that is more compact, ``hotter'' and less luminous in
gravitational waves.

Our simulations have also shown that the spatial distribution and
composition of the ejected matter vary with spin. Most notably, we have
demonstrated that binaries with high aligned spins eject extremely
neutron-rich material and at lower average velocities. The contrary is
true for binaries with large misaligned spins. Finally, we have
highlighted how neglecting the effects of high spin on the dynamical
ejecta overestimates the amount of ejected mass and can in principle lead
to significant deviations in the modelling of the resulting kilonova
emission. 
This effect will be particularly important for high-mass systems that
either lead to a prompt or delayed collapse within $\lesssim\,10\,\rm
ms$, for which light remnant disks are be obtained
\citep{Rezzolla:2010,Ruiz2017a} and secular mass ejection is subdominant.
By measuring the peak time and magnitude of the luminosity of the red
component compared to the blue component of the kilonova, which in this
case will be dominated by the dynamical ejecta, it should in
principle be possible to infer constraints on the spin of the system
that are complementary to those obtained currently from the
gravitational-wave emission during the inspiral, although modelling
uncertainties in kilonova afterglow modelling still remain large.
To illustrate this, we follow \citet{Abbott2017c}, and in 
the case of a soft EOS such as $\rm TNTYST$, find a difference 
in the peak time and luminosity in kilonova emission of
$t_{p}^{\downarrow\downarrow}/t_{p}^{\uparrow\uparrow}\approx2.2$, and
$L_{p}^{\downarrow\downarrow}/L_{p}^{\uparrow\uparrow}\approx4.3$,
using the values obtained from the simulations in this work.

Our simulations inevitably cover only a small but representative portion
of the space of parameters. Further studies, and the development of new
temperature-dependent EOSs, will be needed to set tighter constraints on
how the dynamically ejected matter relates to the stiffness of the EOS
and to the spin configuration. While our study only addresses equal-mass
systems, the precise interplay of spin and unequal-mass systems remains
to be explored. Nevertheless, the conclusions drawn in this work apply to
realistic mass ratios compatible with GW170817\footnote{Assuming the low
  spin prior.}  ($q\gtrsim0.8$), or to observed binary systems
\citep{ZhuX2018}, $q\gtrsim0.9$. Only when extreme mass ratios are
considered ($q\lesssim 0.7$), spin effects could be subdominant to
tidally driven mass ejection \citep{Dietrich2017c}, although the
numerical error budget of these simulations remains very
large. Additionally, when considering equal mass misaligned systems
$\left( \downarrow\downarrow \right)$, we have found an increase in mass
ejection similar to the one expected for moderate mass ratios
$q\gtrsim0.9$, highlighting a degeneracy between misaligned spinning
systems and unequal-mass binaries. By incorporating finite-temperature
and neutrino effects we were also able to show that the distribution of
outflow is more shock heated and hence expected to be more isotropic than
tidally driven mass ejection in unequal mass system. Further studies will
be required to investigate whether the two effects can, hence, be
discerned in observations of red and blue kilononova compontents,
especially for high-mass systems where the secular ejection will be
suppressed.  Lastly, it will also be important to investigate how the
difference in the angular momentum and compactness of the remnant
resulting from inital spins translates to different lifetimes and disk
masses after collapse to a black hole.  Long-term self-consistent
simulations accounting for magnetically and neutrino-driven winds will be
needed to clarify this point.



\section*{Acknowledgements}

We thank Antonios Nathanail for useful discussions. Support comes in
part from HGS-HIRe for FAIR; the LOEWE-Program in HIC for FAIR;
``PHAROS'', COST Action CA16214 European Union's Horizon 2020 Research
and Innovation Programme (Grant 671698) (call FETHPC-1-2014, project
ExaHyPE); the ERC Synergy Grant ``BlackHoleCam: Imaging the Event Horizon
of Black Holes'' (Grant No. 610058); the National Science Foundation
(NSF) Grant PHY-1662211, and NASA Grant 80NSSC17K0070. The simulations
were performed on SuperMUC at LRZ in Garching, on the
GOETHE-HLR cluster at CSC in Frankfurt, and on the HazelHen cluster at
HLRS in Stuttgart.




\end{document}